\documentclass[10]{article}
\usepackage{fancyhdr}
\usepackage{extramarks}
\usepackage{amsmath}
\usepackage{amsthm}
\usepackage{amsfonts}
\usepackage{siunitx}
\usepackage{tikz}
\usepackage[plain]{algorithm}
\usepackage{algpseudocode}
\usepackage{multirow}
\usepackage{booktabs}
\usepackage{graphicx}
\usepackage{subfigure}
\usepackage[colorlinks,linkcolor=black,anchorcolor=black,citecolor=black,urlcolor=blue]{hyperref}
\usepackage{amsmath,bm}
\usepackage{booktabs}
\usepackage{mathtools}
\usepackage{amssymb}
\usepackage{caption}
\usepackage{capt-of}
\usepackage{mciteplus}
\usepackage{cite}
\usepackage{mathrsfs}
\usepackage[title,titletoc,toc]{appendix}
\usepackage{xr}
\usepackage{parskip}
\usepackage{soul}
\usepackage{textcomp}
\usepackage[colaction]{multicol}
\usepackage[switch]{lineno}
\usepackage{lipsum}
\usepackage{etoolbox}
\usepackage{longtable}
\usepackage{array}
\usepackage{tablefootnote}
\usepackage{ragged2e}
\newcolumntype{C}[1]{>{\centering\arraybackslash}p{#1}}
\usetikzlibrary{automata,positioning}
\usetikzlibrary{automata,positioning}

\title{Dimensionality reduction for k-means clustering of large-scale influenza mutation datasets}
\author{Emilee Walden$^1$, Jiahui Chen$^1$\footnote{Corresponding author. Email: jiahuic@uark.edu}, and Guo-Wei Wei$^{2,3,4}$  \\
$^1$ Department of Mathematical Sciences, \\
University of Arkansas, Fayetteville, AR 72701, USA.\\
$^2$ 
Department of Mathematics, \\
Michigan State University, MI 48824, USA.\\
$^4$ Department of Electrical and Computer Engineering,\\
Michigan State University, MI 48824, USA. \\
$^5$ Department of Biochemistry and Molecular Biology,\\
Michigan State University, MI 48824, USA. \\
}

\begin{document}

\maketitle

\begin{abstract}
Viral mutations pose significant threats to public health by increasing infectivity, strengthening vaccine resistance, and altering disease severity. To track these evolving patterns, agencies like the CDC annually evaluate thousands of virus strains, underscoring the urgent need to understand viral mutagenesis and evolution in depth. In this study, we integrate genomic analysis, clustering, and three leading dimensionality reduction approaches, namely, principal component analysis (PCA), t-distributed stochastic neighbor embedding (t-SNE), and uniform manifold approximation and projection (UMAP)—to investigate the effects of COVID-19 on influenza virus propagation. By applying these methods to extensive pre- and post-pandemic influenza datasets, we reveal how selective pressures during the pandemic have influenced the diversity of influenza genetics. Our findings indicate that combining robust dimension reduction with clustering yields critical insights into the complex dynamics of viral mutation, informing both future research directions and strategies for public health intervention.
\end{abstract}

\section{Introduction}
Influenza viruses remain a significant global health concern due to their high rates of morbidity and mortality, despite widespread vaccination efforts. Annually, influenza is responsible for approximately 200,000 hospitalizations and 36,000 deaths in the United States alone \cite{thompson2003mortality}, causing over \$3.7 billion in direct medical costs \cite{paget2019global}. Particularly vulnerable populations--such as the elderly, infants, and individuals with chronic conditions -- face especially high mortality risks. 

A key factor contributing to the persistent threat of influenza is the virus's ability to rapidly evolve and evade natural and vaccine-derived immunity, a process known as ``antigenic drift'' \cite{koelle2006epochal,boni2008vaccination,smith2004mapping,bedford2014integrating,bedford2015global}. This continual evolution allows the virus to reinfect individuals who have immunity to previously circulating strains, thus replenishing its pool of susceptible hosts \cite{neher2016prediction,huddleston2020integrating,castro2020silent}. The rapid mutation of influenza viruses poses significant challenges to vaccine development and public health strategies. Understanding the genetic diversity and evolutionary dynamics of influenza viruses is crucial for predicting future outbreaks, designing effective vaccines, and developing targeted treatment strategies. This diversity arises from errors made by viral polymerase during replication within hosts and is shaped by evolutionary forces within and between hosts \cite{mccrone2018genetic}. Population bottlenecks during transmission further influence viral diversity. Analyzing these dynamics provides insights into the mechanisms driving influenza evolution.

Seasonal influenza viruses (type A) evade human immunity through frequent amino acid substitutions in their hemagglutinin (HA) and neuraminidase (NA) surface glycoproteins \cite{hay2001evolution}. To maintain efficacy, vaccines must be regularly updated to match the antigenic properties of circulating strains. The World Health Organization (WHO)’s global influenza surveillance and response system (GISRS) continuously monitors the genotypes and antigenic properties of these viruses, with substantial contributions from WHO Collaborating Centers \cite{barr2014recommendations}. Understanding how new antigenic variants evolve and spread is essential for the selection of effective vaccine strains and for the development of vaccines that are more robust against viral evolution \cite{viboud2020beyond}.
Beyond regular seasonal outbreaks, influenza pandemics have occurred sporadically, with significant outbreaks every 8 to 41 years over several centuries. 
These pandemics can infect up to 50\% of the population in a single year, leading to mortality rates far exceeded those in typical seasons. 
Notable pandemics occurred in 1889, 1918, 1957, and 1968, with the 1918 pandemic being particularly devastating, causing approximately 546,000 excess deaths in the United States and up to 50 million globally \cite{taubenberger2006mm,johnson2002updating}.
In contrast, influenza B viruses, while capable of causing large epidemics, do not cause pandemics, and influenza C viruses typically cause only mild sporadic respiratory illnesses.

Influenza A viruses are the most significant influenza pathogens. A pandemic involving influenza A can occur when a new distinct strain emerges against which people have little to no immunity and which has the capacity to infect and spread efficiently \cite{jeffery20061918,simonsen1999global,morens20071918,johnson2002updating}. These viruses are classified into subtypes based on the HA and NA proteins on their surface, with 18 HA and 11 NA subtypes identified, resulting in over 130 known combinations, mainly in wild birds \cite{obenauer2006large}. The diversity of these viruses is further increased by reassortment, a process where gene segments are exchanged when a host is co-infected by different influenza viruses.
Subtypes such as A (H1N1) and A (H3N2) are particularly prevalent among humans. These subtypes are further divided into numerous genetic clades and sub-clades, adding to the complexity of the virus and posing challenges in vaccine design and effectiveness. Influenza B viruses are divided into two main lineages, B/Yamagata and B/Victoria, which vary geographically and seasonally, complicating vaccine composition and public health responses \cite{paul2013burden}.

Analyzing influenza virus data is crucial to understanding the dynamics of virus evolution, spread, and control. Traditional analytical methods, such as genetic sequencing, approaches based on phylogenetic trees \cite{page2012space,hozumi2024revealing}, and clustering techniques, i.e., $K$-means, are commonly used. However, these unsupervised machine learning techniques, where ground truth is unavailable, face significant limitations when handling the complexity and high dimensionality of influenza data. 
Phylogenetic analysis, a standard method of understanding mutational trends by clustering mutations to reveal evolutionary patterns and transmission pathways, becomes computationally unfeasible as the number of genome samples increases~\cite{page2012space}. This limitation makes it unsuitable for large genomic datasets, necessitating alternative scalable solutions.
In contrast, $K$-means clustering offers better scalability but often under-performs with small sample sizes or in large feature spaces. The method relies on computing distances between cluster centers and individual samples, a process that becomes computationally expensive and memory intensive as the data dimensionality increases. 
The Jaccard distance is commonly used to compare genome sequences \cite{zhou2008approach} because it effectively captures phylogenetic or topological differences between samples. However, a trade-off of using the Jaccard distance is that its feature dimension equals the number of samples. This means that for large sample sizes, such as the 26,696 influenza genome sequences collected from 2009 to 2024, the feature space becomes exceedingly large. This high dimensionality leads to expensive computations, substantial memory requirements, and poor clustering performance. To mitigate these issues, dimensionality reduction techniques are employed to simplify the data prior to clustering \cite{hozumi2021umap}. 

These techniques reduce the number of features while preserving essential structural information, thereby improving computational efficiency and clustering performance. Dimensionality reduction algorithms generally focus on two aspects: (1) Preserving global pairwise distances, and (2) preserving local distances over global distances. The former aims to maintain the overall distance relationships among all data samples. Techniques such as principal component analysis (PCA) \cite{jolliffe2016principal}, Sammon Mapping \cite{sammon1969nonlinear}, and multidimensional scaling (MDS) \cite{chen2008multidimensional} fall into this group. PCA, for instance, reduces dimensions by projecting them onto a new subspace with preserved variances, though it may lose significant information if the number of principal components is not optimally selected. While PCA is adept at uncovering large-scale structures in data, it performs poorly with nonlinear relationships, leading to the adoption of more sophisticated nonlinear methods.  
The latter category aims to maintain the relationships between neighboring data points, which is crucial to capture the intrinsic structure of complex datasets. Techniques include t-distributed stochastic neighbor embedding (t-SNE) \cite{linderman2019fast,van2008visualizing}, uniform manifold approximation and projection (UMAP) \cite{becht2019dimensionality,mcinnes2018umap}, Laplacian eigenmaps \cite{belkin2001laplacian}, and LargeVis \cite{tang2016visualizing}. t-SNE minimizes the divergence between probability distributions representing high- and low-dimensional spaces to preserve local structures but can be computationally intensive and sensitive to hyperparameters, affecting scalability. UMAP builds upon the mathematical foundations of Laplacian eigenmaps and Riemannian geometry to provide a faster and more effective solution. It preserves both local and some global structures by optimizing data layout through fuzzy set cross-entropy loss minimization. A comparison of PCA, UMAP, and t-SNE was performed in the analysis of single-cell RNA-sequence data \cite{cottrell2024k}, indicating PCA's advantages. The performance of PCA can be significantly improved by topological PCA (tPCA) \cite{cottrell2024k}. Furthermore, correlated clustering and projection (CCP)-assisted UMAP and t-SNE were developed to considerably improve the accuracy and robustness of UMAP and t-SNE in their dimensionality reduction and visualization \cite{hozumi2024analyzing}.  

Recent generative models such as variational autoencoders (VAEs) \cite{kingma2013auto} and generative adversarial networks (GANs)~\cite{goodfellow2020generative}, along with their variants, also offer dimensionality reduction capabilities. These models provide latent representations that respect local structures and ensure smooth transitions between data points, achieved through a generative modeling approach rather than directly optimizing local distances. By applying these dimensionality reduction techniques, high-dimensional influenza genomic data become more tractable for clustering algorithms like $K$-means. This combination enables more efficient and effective analysis of viral evolution, spread, and control, addressing the computational and scalability issues inherent in handling large-scale and  high-dimensional datasets.

In this work, our objective is to develop a dimensionality reduction-assisted clustering method to enhance the analysis of large volumes of influenza genome sequences. By comparing PCA, t-SNE, and UMAP in conjunction with K-means clustering, we evaluated their effectiveness and accuracy in extracting meaningful patterns from complex biological data. Our evaluation involves recasting supervised classification problems into K-means clustering scenarios to quantitatively measure the performance and accuracy of each method, thus determining the optimal approach for handling large-scale influenza data.

\section{Results}
\begin{figure}[htb]
    \centering
    \includegraphics[width=0.6\textwidth]{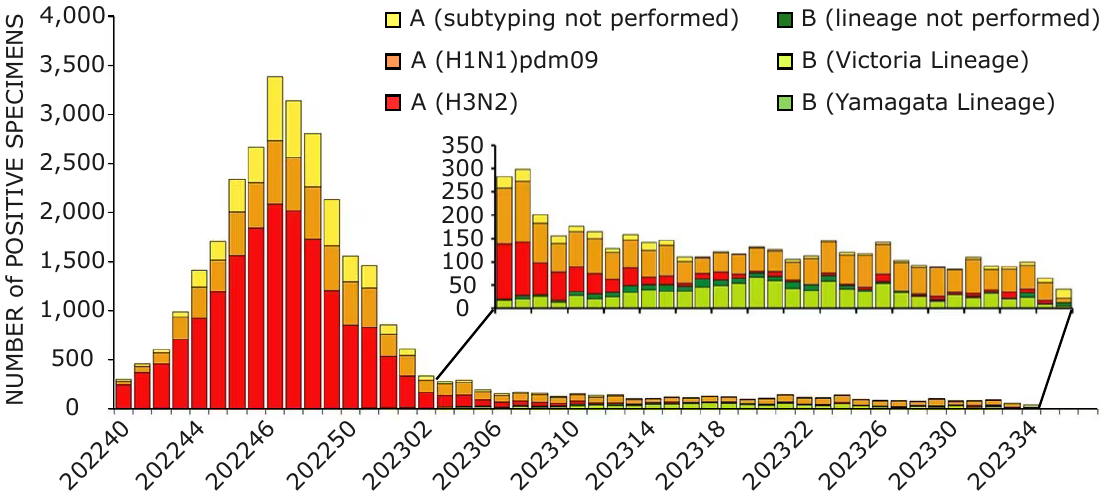}
    \caption{Reported positive tests for influenza by U.S. Public Health Laboratories, summarized nationally from October 2022 to August 2023. Data source: CDC.\cite{CDC} The x-axis represents the year and the corresponding epidemiological week (e.g., “202240” refers to the 40th week of the year 2022).}
    \label{fig:CDC}
\end{figure}

We gather data from NCBI for H3N2 influenza virus sequences from 2009 to 2024, and the total number of samples is 26,696. All sequences included in this study were collected from cases reported in the United States to ensure consistency in regional epidemiological trends. We first get the SNP information by applying the multiple sequence alignment. Next, we calculate the pairwise Jaccard distance of our dataset in order to generate the Jaccard distance-based features. Here, the number of rows is the number of samples (26,696), and the number of columns is the feature size (26,696). The Jaccard distance-based feature is a square matrix. However, due to the large size of samples and features, applying $K$-means clustering directly on the feature of the size of $26,696 \times 26,696$ is a very time-consuming process. To address this issue, we illustrate the PCA, t-SNE, and UMAP methods to reduce the feature size for clustering. According to Figure~\ref{fig:CDC}, H3N2 and H1N1 are the top two subvariants of influenza viruses in terms of reported cases, making them ideal candidates for studying evolutionary trends and genetic diversity. {We also compare the clustering with time by analyzing how the clustering results evolve over different time periods within our dataset.} Specifically, we separate the dataset into two distinct time periods: pre-COVID-19 (2009 to 2019) and post-COVID-19 (2020 to 2024). We then apply the same dimensionality reduction techniques (PCA, t-SNE, UMAP) and $K$-means clustering to each time period. This comparison allows us to assess how the COVID-19 pandemic may have impacted the evolution and genetic diversity of the H3N2 and H1N1 influenza viruses. By analyzing these temporal subsets, we aim to identify any significant shifts in clustering patterns or the emergence of new clusters in the post-COVID-19 period. The time-based comparison will help us understand if there is a significant temporal trend or shift in genetic diversity within the virus over time.

Figure~\ref{fig:CDC} illustrates the number of positive influenza specimens, categorized by various virus subtypes, from a specific time frame (2022-2023) of the CDC~\cite{CDC}. It highlights the dominance of the H3N2 strain of the influenza A virus in circulating flu cases. The H1N1 strain ranks second after H3N2, making it another variant of concern due to its substantial contribution to influenza cases. These two strains have been significant contributors to seasonal flu outbreaks, with their impact being noticeable due to their high transmission rates compared to other influenza subtypes. The seasonal impacts are evident in the variation of positive specimen numbers over time, with a sharp peak occurring during the colder months. The decline following the peak reflects the typical seasonal pattern of influenza transmission. Additionally, the figure includes an inset that provides a more detailed view of a lower but sustained level of influenza activity in later months, indicating that while flu cases drop post-peak, the virus continues to circulate at lower levels. The presence of influenza B cases, though relatively lower in number, also contributes to the overall burden of flu, with subtypes such as the Victoria and Yamagata lineages being detected. This seasonal variation underscores the cyclical nature of flu outbreaks and the importance of tracking viral subtypes, particularly H3N2 and H1N1, which have shown a capacity for widespread transmission. The dataset captures the dynamic nature of flu activity, influenced by environmental conditions and other factors, including the overlapping spread of respiratory illnesses such as COVID-19 in recent years.

\subsection{Dimension reduction analysis of H3N2 evolution}
As H3N2 is the dominant strain of influenza, we focused our analyses on H3N2 sequences collected from 2009 to 2024. We employed three dimensionality reduction methods, PCA, t-SNE, and UMAP, to cluster the pre-COVID-19 dataset (2009–2019) and then projected the post-COVID-19 dataset (2020–2024) on the resulting pre-pandemic clustering. This approach enabled us to examine genetic changes of the H3N2 virus before and after the COVID-19 pandemic.

\begin{figure}[htb]
    \centering
    \includegraphics[width=\textwidth]{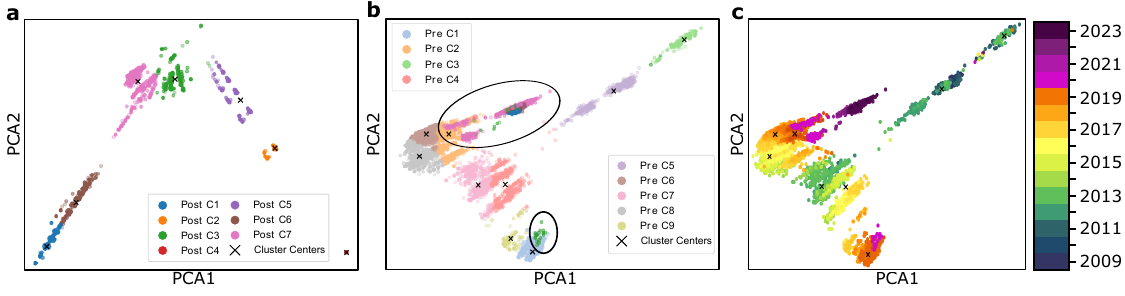}
    \caption{{\bf a}. Clustering of the dataset post-COVID-19, demonstrating the new clustering structure. {\bf b}. PCA clustering before COVID-19, showing 9 clusters, with two highlighted circles representing the post-COVID-19 dataset. {\bf c}. Time-involved evolution of the entire dataset from 2009 to 2024 from b.}
    \label{fig:PCA_H3N2}
\end{figure} 

\begin{figure}[htb]
    \centering
    \includegraphics[width=\textwidth]{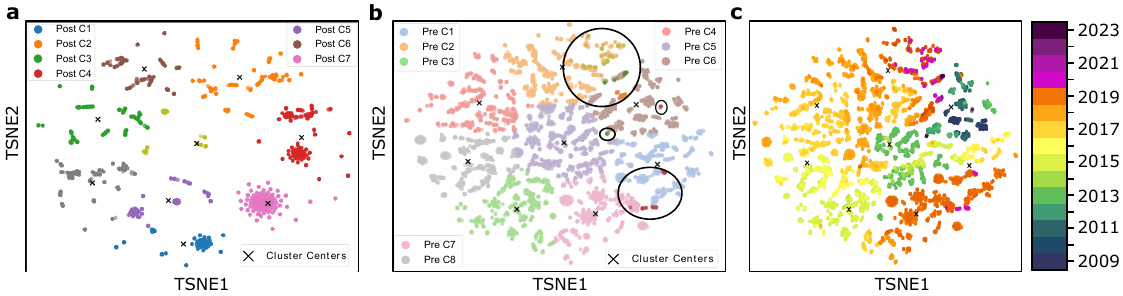}
    \caption{{\bf a}. Clustering of the dataset post-COVID-19, demonstrating the new clustering structure. {\bf b}. t-SNE clustering before COVID-19, showing 8 clusters, with four highlighted circles representing the post-COVID dataset. {\bf c}. Time-involved evolution of the entire dataset from 2009 to 2024 from b.}
    \label{fig:TSNE_H3N2}
\end{figure} 

\begin{figure}[htb]
    \centering
    \includegraphics[width=\textwidth]{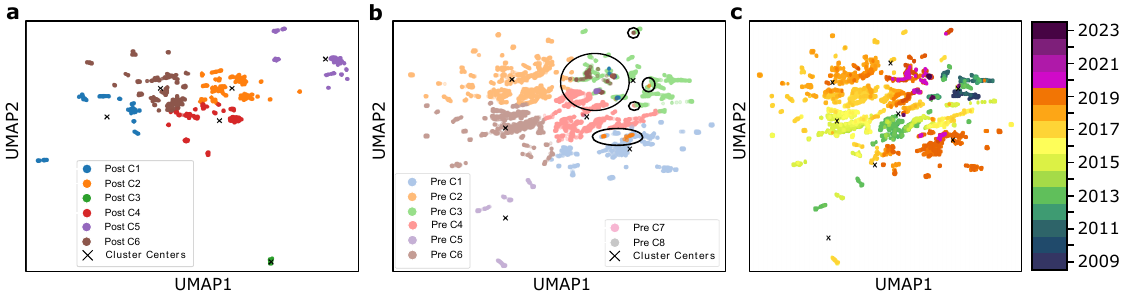}
    \caption{{\bf a}. Clustering of the dataset post-COVID-19, demonstrating the new clustering structure. {\bf b}. UMAP clustering before COVID-19, showing 8 clusters, with four highlighted circles representing the post-COVID dataset.  {\bf c}. Time-involved evolution of the entire dataset from 2009 to 2024 from b.}
    \label{fig:UMAP_H3N2}
\end{figure} 

Figures \ref{fig:PCA_H3N2}, \ref{fig:TSNE_H3N2}, and \ref{fig:UMAP_H3N2} each illustrate PCA, t-SNE, and UMAP clustering results, respectively, with panels ({\bf a}) showing the post-COVID-19 dataset (2020–2024), panels ({\bf b}) depicting the pre-COVID-19 dataset (2009–2019) along with highlighted regions that represent the projected locations of post-COVID-19 samples, and panels ({\bf c}) visualizing the entire time span from 2009 to 2024 colored by year of collection. In Figure \ref{fig:PCA_H3N2} (PCA), the newly formed clusters in panel ({\bf a}) contrast with multiple well-defined pre-COVID-19 clusters in panel ({\bf b}), which reveals where 2020–2024 sequences map onto the 2009–2019 lineage; panel ({\bf c}) offers a same timeline in panel ({\bf b}) of genetic evolution. Figure \ref{fig:TSNE_H3N2} (t-SNE) similarly shows clear segmentation of post-2020 samples in panel ({\bf c}) offers a timeline of genetic evolution. Figure \ref{fig:TSNE_H3N2} (t-SNE) similarly shows clear segmentation of post-2020 samples in panel ({\bf a}), which panel ({\bf b}) reveals fewer pre-COVID-19 clusters in general but distinct regions where post-pandemic data appear, and panel ({\bf c}) capturing the gradual progression of H3N2 across years. Lastly, Figure \ref{fig:UMAP_H3N2} (UMAP) depicts in panel ({\bf a}) the pronounced new clusters emerging post-COVID-19, in panel ({\bf b}) the eight main groups among the pre-2020 data overlaid with several post-2020 clusters, and in panel ({\bf c}) the year-by-year distribution from 2009 through 2024, underscoring changing patterns in H3N2's genetic structure.

\noindent
{\bf Pre- and post-COVID-19 clustering.}
In all three dimensionality reduction approaches (Figures \ref{fig:PCA_H3N2}-\ref{fig:UMAP_H3N2}), the pre-COVID-19 data (2009-2019) group into well-defined clusters, illustrating diverse genetic lineages of H3N2 before 2020. Projecting the post-COVID-19 data (2020-2024) onto these pre-pandemic clusters reveals that some of the recent strains align closely with an existing cluster, while the majority diverge into newly formed groups. This pattern suggests that although a fraction of H3N2 strains continued along older evolutionary paths, most sequences collected after the pandemic display significant genetic deviation, likely reflecting selective pressures or shifts in viral spread.

\noindent
{\bf Detailed observation in t-SNE and UMAP.}
While PCA identifies nine pre-pandemic clusters, both t-SNE and UMAP characterize eight. Under t-SNE (Figure \ref{fig:TSNE_H3N2}), the post-COVID-19 dataset projects into four major groups. Notably, a red cluster separates from the main cloud in Figure \ref{fig:TSNE_H3N2}{\bf a}, and three small isolated areas in Figure \ref{fig:TSNE_H3N2}{\bf b} indicate newly emerging variants (as further confirmed by the time-colored plot in Figure \ref{fig:TSNE_H3N2}{\bf c}. In UMAP (Figure \ref{fig:UMAP_H3N2}), the post-pandemic data similarly group into five main clusters, with one orange cluster deviating markedly from the rest, and three small, isolated regions corresponding to the most recent data (shown in Figure \ref{fig:UMAP_H3N2}{\bf c}). These consistent results across two different nonlinear methods highlight the robustness of the observed shifts in H3N2's genetic structure.

\noindent
{\bf Temporal evolution.}
The panels labeled ({\bf c}) in Figures \ref{fig:PCA_H3N2}-\ref{fig:UMAP_H3N2} collectively offer a year-by-year view of H3N2 evolution, from 2009 through 2024. Before the COVID-19 pandemic, clusters remain relatively stable with moderate genetic progression. After 2020, the emergence of new clusters and changes in existing ones become more pronounced, indicating a surge in genetic diversification in PCA analysis. This observation underscores how the COVID-19 era coincided with substantial modifications in H3N2's evolutionary trajectory, potentially influenced by altered epidemiological conditions and shifts in human behavior during the pandemic.

\subsection{Dimension reduction analysis of H1N1 evolution}
\begin{figure}[H]
    \centering
    \includegraphics[width=\textwidth]{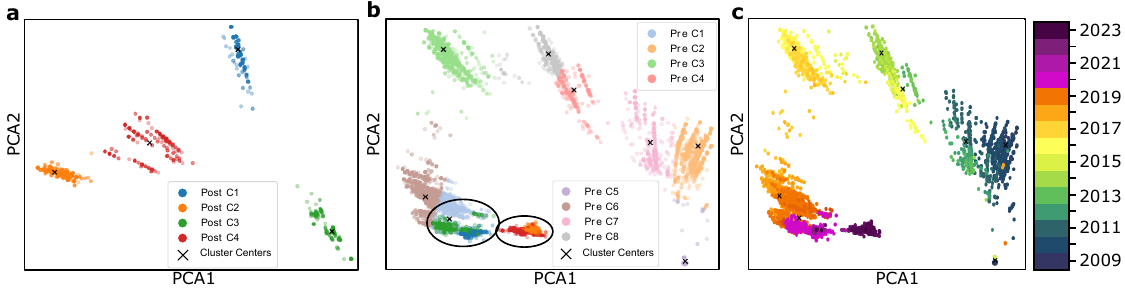}
    \caption{{\bf a}. Clustering of the dataset post-COVID-19, demonstrating the new clustering structure. {\bf b}. PCA clustering before COVID-19, showing 8 clusters, with two highlighted circles representing the post-COVID dataset. {\bf c}. Time-involved evolution of the entire dataset from 2009 to 2024 from b.}
    \label{fig:PCA_H1N1}
\end{figure}

\begin{figure}[H]
    \centering
    \includegraphics[width=\textwidth]{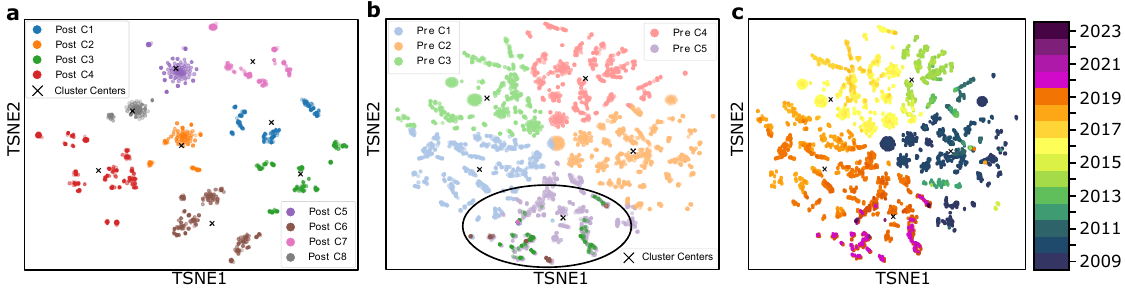}
    \caption{{\bf a}. Clustering of the dataset post-COVID-19, demonstrating the new clustering structure. {\bf b}. t-SNE clustering before COVID-19, showing 5 clusters, with one highlighted circle representing the post-COVID dataset. {\bf c}. Time-involved evolution of the entire dataset from 2009 to 2024 from b.}
    \label{fig:TSNE_H1N1}
\end{figure}

\begin{figure}[H]
    \centering
    \includegraphics[width=\textwidth]{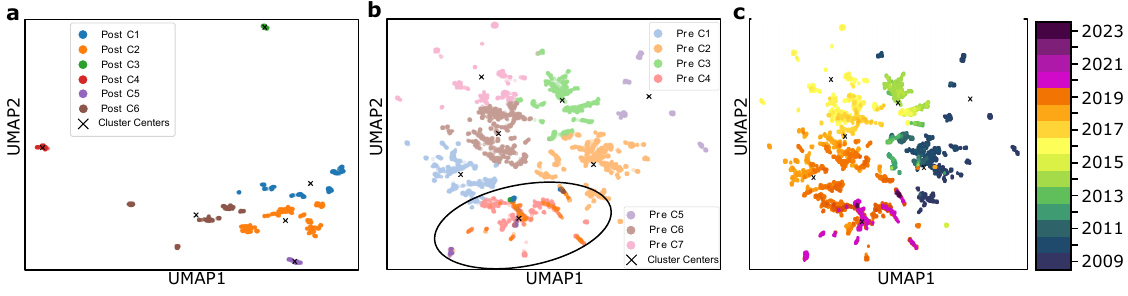}
    \caption{{\bf a}. Clustering of the dataset post-COVID-19, demonstrating the new clustering structure. {\bf b}. UMAP clustering before COVID-19, showing 7 clusters, with one highlighted circle representing the post-COVID dataset. {\bf c}. Time-involved evolution of the entire dataset from 2009 to 2024 from b.}
    \label{fig:UMAP_H1N1}
\end{figure}

As the second most prevalent influenza subvariant after H3N2, H1N1 also underwent a comparable dimension reduction analysis using PCA, t-SNE, and UMAP. Figures~\ref{fig:PCA_H1N1}–\ref{fig:UMAP_H1N1} show the results for the post-COVID-19 data (2020–2024) and the pre-COVID-19 data (2009–2019), highlighting both newly formed clusters and how the recent samples project onto the older lineages. This approach once again helps capture possible shifts in viral evolution that may have coincided with the COVID-19 era.

Figures \ref{fig:PCA_H1N1}, \ref{fig:TSNE_H1N1}, and \ref{fig:UMAP_H1N1} illustrate the post-COVID-19 H1N1 dataset in panel ({\bf a}), the pre-COVID-19 data with overlaid post-2020 points in panel ({\bf b}), and a year-by-year view from 2009 to 2024 in panel ({\bf c}). Similar to H3N2, PCA tends to isolate more clusters, while t-SNE and UMAP produce fewer but still consistent groupings, highlighting how the most recent data diverge from earlier H1N1 strains.

\noindent
{\bf Clustering Pre- and post-COVID-19.}
Similar to H3N2, the pre-pandemic H1N1 data are well structured under each dimensional reduction method, but PCA appears to produce more (and clearer) clusters than t-SNE or UMAP. For instance, in Figure~\ref{fig:PCA_H1N1}{\bf b}, two separate clusters arise within the post-2020 data when projected onto pre-2020 clusters, whereas Figures~\ref{fig:TSNE_H1N1}{\bf b} and \ref{fig:UMAP_H1N1}{\bf b} each show only one major group. This discrepancy suggests that PCA may parse out finer distinctions among circulating H1N1 strains, while t-SNE and UMAP group some lineages together.

\noindent
{\bf Observations on recent H1N1 data.}
A key finding is that the most recent points (2023–2024) tend to lie on the periphery of the main clusters, echoing the ``elusiveness'' observed in H3N2 analyses. In Figures~\ref{fig:TSNE_H1N1}{\bf a} and \ref{fig:UMAP_H1N1}{\bf a}, these points are more dispersed, implying ongoing or accelerated genetic changes. Figure~\ref{fig:PCA_H1N1}{\bf c} confirms that some of the newest H1N1 samples form a separate cluster altogether, suggesting novel mutations or lineages divergent from established strains. This pattern, i.e., fewer and more diffuse groupings under t-SNE and UMAP, respectively, and more distinct clusters via PCA, parallels the H3N2 results, pointing to possible selective pressures on recent H1N1 evolution.

\section{Discussion}
Our analysis demonstrates that dimensionality reduction techniques, specifically PCA, t-SNE, and UMAP, when combined with $K$-means clustering, effectively handle large-scale influenza genome datasets. By applying these methods to the Jaccard distance-transformed influenza virus sequences from 2009 to 2024, we discovered meaningful patterns that reflect the genetic drift of the virus over time.

\noindent
{\bf Genetic drift captured by PCA.} PCA effectively captured the genetic drift in influenza virus data, whose projection of the 2020-2024 dataset in Figures~\ref{fig:PCA_H3N2} and \ref{fig:PCA_H1N1} revealed a split into two distinct groups, showing a linear progression consistent with the accumulation of mutations over time. One trajectory includes the most recent data moving further away from previous clusters, indicating the emergence of novel virus strains. The other trajectory contains data primarily from around 2020, with no new data closer to 2024, suggesting that these virus strains may have ceased spreading. This linear separation aligns with the expected patterns of genetic drift, where mutations accumulate gradually, leading to divergence from ancestral strains. The ability of PCA to capture these temporal and evolutionary trends underscores its utility in monitoring the genetic evolution of influenza viruses. 

\noindent
{\bf Nonlinear methods.} In contrast, t-SNE and UMAP, which are nonlinear dimensionality reduction techniques that preserve local structures, provided additional insights by identifying five clusters within the same dataset. The largest cluster for both t-SNE and UMAP corresponds to the most recent data, similar to PCA's findings. A slightly smaller cluster contains data from around 2020, again mirroring the PCA results. However, t-SNE and UMAP also revealed two intermediate clusters comprising data near 2024 situated between the two larger clusters of H3N2. These additional clusters suggest the presence of emerging or transitional virus strains that may not be as apparent in the linear PCA projection. The nonlinear approaches of t-SNE and UMAP are more sensitive to local relationships in high-dimensional data, allowing them to detect subtle genetic variations and substructures within the influenza virus populations.

\noindent
{\bf Impact of COVID-19 pandemic.} Our results show evidence of the impact of the COVID-19 pandemic on influenza virus diversity. The public health interventions implemented to control COVID-19, such as social distancing, mask wearing, travel restrictions, and lockdowns, likely contributed to a significant reduction in the transmission of influenza viruses during this period \cite{maltezou2020influenza}. This is reflected in our results by the apparent cessation of certain virus strains after 2020, as indicated by the absence of new data in those clusters moving towards 2024. The reduced diversity and spread of influenza subvariants can be attributed to decreased opportunities for viral transmission due to these interventions. Projecting post-COVID-19 influenza data onto the manifold constructed from pre-COVID-19 data offers valuable insights into how pandemic response measures influenced influenza virus evolution. The shifts in clustering patterns suggest alterations in the typical genetic drift trajectory, possibly due to reduced transmission events resulting in to fewer opportunities for mutation and reassortment.

\noindent
{\bf Preservation of temporal information and future research.} 
Overall, employing the Jaccard distance metric combined with dimensionality reduction techniques preserves temporal information essential for both retrospective and prospective influenza virus analysis. PCA effectively captures the broad patterns of genetic drift over time, while t-SNE and UMAP provide detailed insights into local genetic variations and emerging strains. This multi-faceted approach enhances our understanding of influenza virus evolution and highlights the importance of utilizing multiple dimensionality reduction techniques to gain a comprehensive view of complex biological data. Our findings illustrate the utility of scalable analytical methods in monitoring viral evolution, which is crucial for the selection of vaccine strains and anticipating future outbreaks. The ability to detect emerging novel strains and understand the impact of public health interventions on virus evolution can inform strategies for influenza prevention and control. 

Future work could involve implementing recent generative models and deep learning methods to further enhance the analysis of large-scale influenza datasets. Techniques such as variational autoencoders (VAEs) can capture complex, nonlinear relationships within high-dimensional data, potentially improving clustering performance and the detection of subtle genetic variations. In addition, introducing temporal information into time-lagged VAE can provide a more nuanced understanding of viral evolution over time. By using these advanced computational techniques, we can improve the scalability and accuracy of our analyses, ultimately contributing to more effective public health responses.

\section{Conclusion}
In conclusion, our findings illustrate how dimensionality reduction methods, particularly PCA, effectively capture genetic drift in influenza viruses, while nonlinear methods such as t-SNE and UMAP uncover subtler layers of genetic variation. The pronounced effects of the COVID-19 pandemic and accompanying public health measures on the diversity of influenza viruses further illustrate the need for robust analytical approaches. By combining advanced dimensionality reduction and clustering techniques and integrating generative models and temporal data, future research can provide deeper insights into influenza evolution. Such comprehensive strategies will be instrumental in refining surveillance, guiding vaccine development, and improving the overall management of influenza outbreaks.

\section{Methods}
\subsection{Sequence Preparation}
Influenza virus sequences were obtained from NCBI database, selecting for human type A nucleotide sequences from the USA. Sub-types H1N1 and H3N2 from the HA protein were collected with 13,728 and 26,696 complete sequences respectively from  2009 to 2024. Each sequence was aligned using the multiple sequence alignment package Clustal Omega. Each sequence was then shortened to a defined common start and stop index among all sequences given the various sequence lengths before alignment. 
 
\subsection{Single Nucleotide Polymorphisms Position Based Features}
Single Nucleotide Polymorphisms (SNPs) represent variations where a single base pair differs from the reference sequence. 
Given a uniform sequence length $M$, an $N \times M$ difference matrix was computed using the first sequence in the list as the reference, 
where $N$ denotes the number of sequences. The resulting $N \times M$ position-based feature matrix is defined as
\begin{equation}
    S(i,j) = \begin{cases}
        0, & \text{if no SNP at position $j$ in sequence $i$} \\
        \phi (w,m),  
         & \text{if SNP at position $j$ in sequence $i$}
    \end{cases}
\end{equation}
such that $w,m \in \{A,T,G,C,-\}$ and $\phi: S \times S \rightarrow \mathbb{Z}$, so each row represents a sample, with $S(i,j)$ providing a numerical representation of SNP positions. To determine SNP positions, a reference sequence was selected based on majority voting using the first week’s collected data. Mutations were then identified relative to this reference. For Jaccard feature calculations, pairwise Jaccard distances were computed among all sequences based on SNP positions, without considering mutation types. Due to the large number of samples, dimensionality reduction techniques were applied to improve computational efficiency.

\subsection{Jaccard Calculations}
The Jaccard distance measures the dissimilarity between two sets and is widely used in phylogenetic studies of SNP profiles. 
In this work, the Jaccard distance is used to compare SNP profiles of influenza genome samples. 
For two sets $A$ and $B$, the Jaccard distance is defined as
\begin{equation}
    d_J(A,B) = 1 - \frac{|A \cap B|}{|A \cup B|}
\end{equation}
Given $N$ SNP profiles aligned to the reference influenza genome sequence, 
let $S_i$, $i=1,2,\ldots,N$ represent the set of mutation positions in the $i$th sample. The Jaccard distance between two sets $S_i$ and $S_j$ is denoted by $d_J(S_i,S_j)$. By calculating the pairwise distances between all samples, we construct an $N\times N$ Jaccard distance matrix $D$. 
This distance established a metric for the collection of all finite sets~\cite{levandowsky1971distance}.

\subsection{Principal Component Analysis}
Principal Component Analysis (PCA) is a widely used statistical technique for dimensionality reduction, particularly useful in the exploratory analysis of high-dimensional data~\cite{jolliffe2016principal}. The method assumes linear relationships between variables without making specific assumptions about the distribution of the data features. Therefore, PCA is especially effective for analyzing new data and studying linear trends, such as in the evolution of the influenza virus.
PCA transforms a large set of correlated variables into a smaller set of linearly uncorrelated variables known as principal components. These principal components are orthogonal to each other and are ordered so that the first component captures the maximum variance in the data, followed by subsequent components capturing the remaining variance.

Let $\{x_i\}_{i=1}^N$ represent $N$ data points in an $M$-dimensional space, where $M$ is the number of features. The data is organized into an $N \times M$ matrix $X$, with each row representing a data point. PCA seeks to find a linear combination of the columns of $X$ that maximizes the variance:
\begin{equation}
    \sum_{j=1}^n a_jx_j = Xa,
\end{equation}
where $a_1$, $a_2$, $\cdots$, $a_n$ are principal component vectors. The variance of this linear combination is defined as
\begin{equation}
    \text{var}(Xa) = a^T X^TX a,
\end{equation}
where $X^TX$ is the covariance matrix of the dataset. The first principal component, which maximizes the variance, can be computed iteratively using Rayleigh's quotient:
\begin{equation}
    a_1 = \arg \max_a \frac{a^T X^TX a}{a^T a}.
\end{equation}

Subsequent principal components are computed by maximizing the variance of the residual data matrix:
\begin{equation}
    \hat{X}_k = X - \sum_{j=1}^{k-1} Xa_ja_j^T,
\end{equation}
where $k$ represents the $k$th principal component. This process removes the contribution of the first $k-1$ components from the original matrix $X$. The complexity of the method scales with the number of components one seeks to find. In practice, the hope is that the first few components provide a good representation of the original data matrix $X$.

\subsection{t-SNE}
The t-distributed stochastic neighbor embedding (t-SNE) is a nonlinear dimensional reduction algorithm particularly effective for mapping high-dimensional data into two or three-dimensional space.
The algorithm operates in two main stages. 
First, it constructs a probability distribution over pairs of data points in the high-dimensional space, where pairs of nearby points are assigned high probabilities, and pairs of distant points are given low probabilities. Second, t-SNE defines a similar probability distribution in the low-dimensional embedded space and minimizes the Kullback-Leibler (KL) divergence between these two distributions~\cite{van2008visualizing}.

Let $\{x_i\}_{i=1}^N$ represent a high-dimensional dataset, where each $x_i \in \mathbb{R}^M$. 
The t-SNE algorithm first constructs a probability distribution $P$ over pairs of data points in the high-dimensional space. 
The conditional probability $p_{j|i}$ that point $x_i$ would pick $x_j$ as its neighbor is defined as
\begin{equation}
    p_{j|i} = \frac{\exp(-||x_i - x_j||^2 / 2\sigma_i^2)}{\sum_{k \neq i} \exp(-||x_i - x_k||^2 / 2\sigma_i^2)}, i \neq j,
\end{equation}
where $\sigma_i$ is the variance of the Gaussian centered at $x_i$. The variance $\sigma_i$ is determined by the perplexity parameter, which controls the number of nearest neighbors considered in the high-dimensional space. The perplexity is a hyperparameter that is typically set between 5 and 50. The probability distribution $P$ is defined as
\begin{equation}
    p_{ij} = \frac{p_{j|i} + p_{i|j}}{2N},
\end{equation}

In the second step, the algorithm learns a $k$-dimensional embedding of the data points in the low-dimensional space, $\{y_i\}_{i=1}^N$, where $y_i \in \mathbb{R}^k$. 
The probability distribution $Q$ over pairs of data points in the low-dimensional space is defined as
\begin{equation}
    q_{ij} = \frac{(1 + ||y_i - y_j||^2)^{-1}}{\sum_{k \neq l} (1 + ||y_k - y_l||^2)^{-1}}, i \neq j.
\end{equation}
The low dimensional embedding is found by minimizing the KL divergence between the two probability distributions $P$ and $Q$:
\begin{equation}
    C = \sum_i KL(P_i || Q_i) = \sum_i \sum_j p_{ij} \log \frac{p_{ij}}{q_{ij}}. 
\end{equation}

\subsection{UMAP}
Uniform Manifold Approximation and Projection (UMAP) is a nonlinear dimensionality reduction technique that relies on three key assumptions: the data is uniformly distributed on a Riemannian manifold, the Riemannian metric is locally constant, and the manifold is locally connected. Unlike t-SNE, which uses a probabilistic model, UMAP is a graph-based algorithm. The core idea of UMAP is to create a predefined $k$-dimensional weighted graph representation of the original high-dimensional data points, minimizing the edge-wise cross-entropy between the weighted graph and the original data. The $k$-dimensional eigenvectors of the UMAP graph are then used to represent each of the original data points. This section provides a computational view of UMAP; for a more theoretical treatment, readers are referred to Ref.\cite{mcinnes2018umap}.

Like t-SNE, UMAP takes input data $X = \{x_i\}_{i=1}^N$, where each $x_i \in \mathbb{R}^M$, and seeks an optimal low-dimensional representation $Y = \{y_i\}_{i=1}^N$, where $y_i \in \mathbb{R}^k$. The first stage involves the construction of weighted $k$-neighbor graphs. In UMAP, a metric $d: X \times X \rightarrow \mathbb{R}^+$ is defined on the input data $X$. For a given $k$, the $k$-neighbor graph is constructed by connecting each data point $x_i$ to its $k$ nearest neighbors under the metric $d$. For each $x_i$, let
\begin{equation}
    \rho_i = \min\left\{d(x_i, x_j): x_j \in \text{NearestNeighbors}(x_i, k)\right\},
\end{equation}
where $\sigma_i$ is defined by the equation
\begin{equation}
    \sum_{j=1}^k \exp\left(-\frac{\max\left\{d(x_i, x_j) - \rho_i, 0\right\}}{\sigma_i}\right) = \log_2(k),
\end{equation}
ensuring that at least one data point is connected to $x_i$ with an edge weight of 1 and that the expected number of neighbors is $k$. A weighted directed graph $\bar{G} = (V,E,\omega)$ is then defined, where $V$ is the set of vertices (the data $X$), $E$ is the set of edges $E = \{(x_i, x_j)|x_j  \in \text{NearestNeighbors}(x_i, k), 1 \leq i \leq N\}$, and $\omega$ is the edge weight given by
\begin{equation}
    \omega_{ij} = \exp\left(-\frac{\max\left\{d(x_i, x_j) - \rho_i, 0\right\}}{\sigma_i}\right).
\end{equation}

UMAP then symmetrizes the directed graph $\bar{G}$ to define an undirected weighted graph $G$. Let $A$ be the adjacency matrix of $\bar{G}$. The symmetric matrix $B$ is obtained as
\begin{equation}
    B = A + A^T - A \circ A^T,
\end{equation}
where $\circ$ denotes the Hadamard product. UMAP evolves an equivalent weighted graph $H$ in the low-dimensional space $\{y_i\}_{i=1}^N \in \mathbb{R}^k$ by applying attractive and repulsive forces to the data points. The attractive force pulls similar data points closer together, while the repulsive force pushes dissimilar points apart. The objective is to find the optimal low-dimensional coordinates $Y$ that minimize the cross-entropy between the weighted graph $H$ and the original high-dimensional data $X$. The evolution of the UMAP graph Laplacian $G$ can be regarded as a discrete approximation of the Laplace-Beltrami operator on the manifold defined by the data~\cite{mcinnes2018umap}.
However, UMAP may not perform well if the data points are not uniformly distributed. If some data points have $k$ significant neighbors while others have significantly more ($k' \gg k$), the $k$-dimensional UMAP may be inefficient. Currently, no algorithm automatically determines the critical minimum $k_{\text{min}}$ for a given dataset.

\subsection{$K$-means Clustering}
$K$-means clustering is a widely used unsupervised learning algorithm in machine learning that partitions a dataset ${x_i}_{i=1}^N$ into $k$ clusters, ${C_1, C_2, \ldots, C_k}$, where $k \leq N$. The algorithm starts by selecting $k$ centroids, either randomly or by a heuristic method, and then assigns each data point to the nearest centroid. The centroids are then updated iteratively by minimizing the within-cluster sum of squares (WCSS), defined as
\begin{equation}
W(C) = \sum_{j=1}^{k} \sum_{x_i \in C_j} ||x_i - \mu_j||^2,
\end{equation}
where $\mu_j$ is the centroid of cluster $C_j$. This process continues until convergence, where the centroids stabilize, resulting in a locally optimal clustering solution. The method, however, typically finds the optimal centroids for a given number of clusters $k$. In practical applications, determining the optimal number of clusters is also crucial. To identify the best $k$, the elbow method is often used. This method involves plotting the WCSS against the number of clusters and selecting the inflection point on the plot, which indicates the optimal number of clusters.

\subsection{Implementation}
After sequences were aligned and prepared, the difference matrix was calculated from SNP sequences, then the Jaccard distance matrix was found from the difference matrix. 
PCA was then conducted on the Jaccard distance matrix to 2 components. The elbow method was used to find an approximation of the number of clusters for the reduced dimension dataset, and k-means clustering was run using this initial k-value to find the cluster centers. The two columns of PCA were used create a 2D scatter plot of the reduced data to visualize defined clusters.
Similarly, UMAP and t-SNE were each used to reduce the distance matrix to 2 components in each method. The elbow method and k-means clustering were used to define cluster centers, and the two columns of each were used to plot the data in a 2D scatter plot.

\subsection{Transforming Data}
The sequences for 2009-2019 were then separated from the data for 2020-2024. The difference matrix and Jaccard distance matrix for each subset was determined. The sequences for each time range were reduced using PCA, t-SNE, and UMAP, and the clusters were defined. 
The difference matrix was then found of the subset from 2020-2024 using the first sequence of the 2009-2019 dataset as the reference sequence to ensure the integer encoding remains consistent. A Jaccard distance matrix was calculated by comparing each row of similarly encoded 2020-2024 difference matrix against the difference matrix for data from 2009-2019. The transform function for both PCA, UMAP and t-SNE, fit to the data for 2009-2019, was applied to reduce the data for 2020-2024 to the same transformation. 

\section*{Data Availability}
The influenza data used in this study was obtained from the publicly accessible NCBI database. The dataset can be accessed through the following link: \url{https://www.ncbi.nlm.nih.gov/genomes/FLU/Database/nph-select.cgi}.

\section*{Acknowledgment}	
This work was supported in part by the Arkansas biosciences institute seed grant, UofA Honors mentor funds, NIH grants  R01AI164266 and R35GM148196, NSF grants DMS-2052983 and IIS-1900473, and  MSU Research Foundation.


\begin{thebibliography}{10}

\bibitem{thompson2003mortality}
William~W Thompson, David~K Shay, Eric Weintraub, Lynnette Brammer, Nancy Cox,
  Larry~J Anderson, and Keiji Fukuda.
\newblock Mortality associated with influenza and respiratory syncytial virus
  in the united states.
\newblock {\em Jama}, 289(2):179--186, 2003.

\bibitem{paget2019global}
John Paget, Peter Spreeuwenberg, Vivek Charu, Robert~J Taylor, A~Danielle
  Iuliano, Joseph Bresee, Lone Simonsen, Cecile Viboud, et~al.
\newblock Global mortality associated with seasonal influenza epidemics: New
  burden estimates and predictors from the glamor project.
\newblock {\em Journal of global health}, 9(2), 2019.

\bibitem{koelle2006epochal}
Katia Koelle, Sarah Cobey, Bryan Grenfell, and Mercedes Pascual.
\newblock Epochal evolution shapes the phylodynamics of interpandemic influenza
  a (h3n2) in humans.
\newblock {\em Science}, 314(5807):1898--1903, 2006.

\bibitem{boni2008vaccination}
Maciej~F Boni.
\newblock Vaccination and antigenic drift in influenza.
\newblock {\em Vaccine}, 26:C8--C14, 2008.

\bibitem{smith2004mapping}
Derek~J Smith, Alan~S Lapedes, Jan~C De~Jong, Theo~M Bestebroer, Guus~F
  Rimmelzwaan, Albert~DME Osterhaus, and Ron~AM Fouchier.
\newblock Mapping the antigenic and genetic evolution of influenza virus.
\newblock {\em science}, 305(5682):371--376, 2004.

\bibitem{bedford2014integrating}
Trevor Bedford, Marc~A Suchard, Philippe Lemey, Gytis Dudas, Victoria Gregory,
  Alan~J Hay, John~W McCauley, Colin~A Russell, Derek~J Smith, and Andrew
  Rambaut.
\newblock Integrating influenza antigenic dynamics with molecular evolution.
\newblock {\em elife}, 3:e01914, 2014.

\bibitem{bedford2015global}
Trevor Bedford, Steven Riley, Ian~G Barr, Shobha Broor, Mandeep Chadha, Nancy~J
  Cox, Rodney~S Daniels, C~Palani Gunasekaran, Aeron~C Hurt, Anne Kelso, et~al.
\newblock Global circulation patterns of seasonal influenza viruses vary with
  antigenic drift.
\newblock {\em Nature}, 523(7559):217--220, 2015.

\bibitem{neher2016prediction}
Richard~A Neher, Trevor Bedford, Rodney~S Daniels, Colin~A Russell, and Boris~I
  Shraiman.
\newblock Prediction, dynamics, and visualization of antigenic phenotypes of
  seasonal influenza viruses.
\newblock {\em Proceedings of the National Academy of Sciences},
  113(12):E1701--E1709, 2016.

\bibitem{huddleston2020integrating}
John Huddleston, John~R Barnes, Thomas Rowe, Xiyan Xu, Rebecca Kondor, David~E
  Wentworth, Lynne Whittaker, Burcu Ermetal, Rodney~Stuart Daniels, John~W
  McCauley, et~al.
\newblock Integrating genotypes and phenotypes improves long-term forecasts of
  seasonal influenza a/h3n2 evolution.
\newblock {\em Elife}, 9:e60067, 2020.

\bibitem{castro2020silent}
Italo~A Castro, Daniel~MM Jorge, Lucas~M Ferreri, Ronaldo~B Martins, Marjorie~C
  Pontelli, Bruna~LS Jesus, Ricardo~S Cardoso, Miria~F Criado, Lucas Carenzi,
  Fabiana~CP Valera, et~al.
\newblock Silent infection of b and cd8+ t lymphocytes by influenza a virus in
  children with tonsillar hypertrophy.
\newblock {\em Journal of virology}, 94(9):10--1128, 2020.

\bibitem{mccrone2018genetic}
John~T McCrone and Adam~S Lauring.
\newblock Genetic bottlenecks in intraspecies virus transmission.
\newblock {\em Current opinion in virology}, 28:20--25, 2018.

\bibitem{hay2001evolution}
Alan~J Hay, Victoria Gregory, Alan~R Douglas, and Yi~Pu Lin.
\newblock The evolution of human influenza viruses.
\newblock {\em Philosophical Transactions of the Royal Society of London.
  Series B}, 356(1416):1861, 2001.

\bibitem{barr2014recommendations}
Ian~G Barr, Colin Russell, Terry~G Besselaar, Nancy~J Cox, Rod~S Daniels, Ruben
  Donis, Othmar~G Engelhardt, Gary Grohmann, Shigeyuki Itamura, Anne Kelso,
  et~al.
\newblock Who recommendations for the viruses used in the 2013--2014 northern
  hemisphere influenza vaccine: Epidemiology, antigenic and genetic
  characteristics of influenza a (h1n1) pdm09, a (h3n2) and b influenza viruses
  collected from october 2012 to january 2013.
\newblock {\em Vaccine}, 32(37):4713--4725, 2014.

\bibitem{viboud2020beyond}
C{\'e}cile Viboud, Katelyn Gostic, Martha~I Nelson, Graeme~E Price, Amanda
  Perofsky, Kaiyuan Sun, N{\'\i}dia Sequeira~Trov{\~a}o, Benjamin~J Cowling,
  Suzanne~L Epstein, and David~J Spiro.
\newblock Beyond clinical trials: Evolutionary and epidemiological
  considerations for development of a universal influenza vaccine.
\newblock {\em PLoS Pathogens}, 16(9):e1008583, 2020.

\bibitem{taubenberger2006mm}
J~Taubenberger.
\newblock Mm, 1918 influenza: the mother of all pandemics, 2006.

\bibitem{johnson2002updating}
Niall~PAS Johnson and Juergen Mueller.
\newblock Updating the accounts: global mortality of the 1918-1920" spanish"
  influenza pandemic.
\newblock {\em Bulletin of the History of Medicine}, pages 105--115, 2002.

\bibitem{jeffery20061918}
K~Taubenberger Jeffery and Morens David.
\newblock 1918 influenza: The mother of all pandemics.
\newblock {\em Emerging Infectious Diseases}, 12(1):15--22, 2006.

\bibitem{simonsen1999global}
Lone Simonsen.
\newblock The global impact of influenza on morbidity and mortality.
\newblock {\em Vaccine}, 17:S3--S10, 1999.

\bibitem{morens20071918}
David~M Morens and Anthony~S Fauci.
\newblock The 1918 influenza pandemic: insights for the 21st century.
\newblock {\em The Journal of infectious diseases}, 195(7):1018--1028, 2007.

\bibitem{obenauer2006large}
John~C Obenauer, Jackie Denson, Perdeep~K Mehta, Xiaoping Su, Suraj Mukatira,
  David~B Finkelstein, Xiequn Xu, Jinhua Wang, Jing Ma, Yiping Fan, et~al.
\newblock Large-scale sequence analysis of avian influenza isolates.
\newblock {\em science}, 311(5767):1576--1580, 2006.

\bibitem{paul2013burden}
W~Paul~Glezen, Jordana~K Schmier, Carrie~M Kuehn, Kellie~J Ryan, and John
  Oxford.
\newblock The burden of influenza b: a structured literature review.
\newblock {\em American journal of public health}, 103(3):e43--e51, 2013.

\bibitem{page2012space}
Roderic~DM Page.
\newblock Space, time, form: viewing the tree of life.
\newblock {\em Trends in ecology \& evolution}, 27(2):113--120, 2012.

\bibitem{hozumi2024revealing}
Yuta Hozumi and Guo-Wei Wei.
\newblock Revealing the shape of genome space via k-mer topology.
\newblock {\em arXiv preprint arXiv:2412.20202}, 2024.

\bibitem{zhou2008approach}
Tingting Zhou, Keith~CC Chan, Yi~Pan, and Zhenghua Wang.
\newblock An approach for determining evolutionary distance in network-based
  phylogenetic analysis.
\newblock In {\em Bioinformatics Research and Applications: Fourth
  International Symposium, ISBRA 2008, Atlanta, GA, USA, May 6-9, 2008.
  Proceedings 4}, pages 38--49. Springer, 2008.

\bibitem{hozumi2021umap}
Yuta Hozumi, Rui Wang, Changchuan Yin, and Guo-Wei Wei.
\newblock Umap-assisted k-means clustering of large-scale sars-cov-2 mutation
  datasets.
\newblock {\em Computers in biology and medicine}, 131:104264, 2021.

\bibitem{jolliffe2016principal}
Ian~T Jolliffe and Jorge Cadima.
\newblock Principal component analysis: a review and recent developments.
\newblock {\em Philosophical transactions of the royal society A: Mathematical,
  Physical and Engineering Sciences}, 374(2065):20150202, 2016.

\bibitem{sammon1969nonlinear}
John~W Sammon.
\newblock A nonlinear mapping for data structure analysis.
\newblock {\em IEEE Transactions on computers}, 100(5):401--409, 1969.

\bibitem{chen2008multidimensional}
Chun-houh Chen, Wolfgang H{\"a}rdle, Antony Unwin, Michael~AA Cox, and Trevor~F
  Cox.
\newblock {\em Multidimensional scaling}.
\newblock Springer, 2008.

\bibitem{linderman2019fast}
George~C Linderman, Manas Rachh, Jeremy~G Hoskins, Stefan Steinerberger, and
  Yuval Kluger.
\newblock Fast interpolation-based t-sne for improved visualization of
  single-cell rna-seq data.
\newblock {\em Nature methods}, 16(3):243--245, 2019.

\bibitem{van2008visualizing}
Laurens Van~der Maaten and Geoffrey Hinton.
\newblock Visualizing data using t-sne.
\newblock {\em Journal of machine learning research}, 9(11), 2008.

\bibitem{becht2019dimensionality}
Etienne Becht, Leland McInnes, John Healy, Charles-Antoine Dutertre,
  Immanuel~WH Kwok, Lai~Guan Ng, Florent Ginhoux, and Evan~W Newell.
\newblock Dimensionality reduction for visualizing single-cell data using umap.
\newblock {\em Nature biotechnology}, 37(1):38--44, 2019.

\bibitem{mcinnes2018umap}
Leland McInnes, John Healy, and James Melville.
\newblock Umap: Uniform manifold approximation and projection for dimension
  reduction.
\newblock {\em arXiv preprint arXiv:1802.03426}, 2018.

\bibitem{belkin2001laplacian}
Mikhail Belkin and Partha Niyogi.
\newblock Laplacian eigenmaps and spectral techniques for embedding and
  clustering.
\newblock {\em Advances in neural information processing systems}, 14, 2001.

\bibitem{tang2016visualizing}
Jian Tang, Jingzhou Liu, Ming Zhang, and Qiaozhu Mei.
\newblock Visualizing large-scale and high-dimensional data.
\newblock In {\em Proceedings of the 25th international conference on world
  wide web}, pages 287--297, 2016.

\bibitem{cottrell2024k}
Sean Cottrell, Yuta Hozumi, and Guo-Wei Wei.
\newblock K-nearest-neighbors induced topological pca for single cell
  rna-sequence data analysis.
\newblock {\em Computers in biology and medicine}, 175:108497, 2024.

\bibitem{hozumi2024analyzing}
Yuta Hozumi and Guo-Wei Wei.
\newblock Analyzing scrna-seq data by ccp-assisted umap and tsne.
\newblock {\em PloS one}, 19(12):e0311791, 2024.

\bibitem{kingma2013auto}
Diederik~P Kingma.
\newblock Auto-encoding variational bayes.
\newblock {\em arXiv preprint arXiv:1312.6114}, 2013.

\bibitem{goodfellow2020generative}
Ian Goodfellow, Jean Pouget-Abadie, Mehdi Mirza, Bing Xu, David Warde-Farley,
  Sherjil Ozair, Aaron Courville, and Yoshua Bengio.
\newblock Generative adversarial networks.
\newblock {\em Communications of the ACM}, 63(11):139--144, 2020.

\bibitem{CDC}
Cdc yearly lab work on flu viruses infographic.
\newblock
  \url{https://www.cdc.gov/flu/resource-center/freeresources/graphics/infographic-lab-work.htm}.

\bibitem{maltezou2020influenza}
Helena~C Maltezou, Kalliopi Theodoridou, and Gregory Poland.
\newblock Influenza immunization and covid-19.
\newblock {\em Vaccine}, 38(39):6078, 2020.

\bibitem{levandowsky1971distance}
Michael Levandowsky and David Winter.
\newblock Distance between sets.
\newblock {\em Nature}, 234(5323):34--35, 1971.

\end{thebibliography}
\end{document}